# Thickness tunable quantum interference between surface phonon and Dirac plasmon states in thin-films of the topological insulator Bi$_2$Se$_3$

Yuri D. Glinka,[1,2*] Sercan Babakiray,[1] Trent A. Johnson,[1] David Lederman[1]

[1]*Department of Physics and Astronomy, West Virginia University, Morgantown, WV 26506-6315, USA*
[2]*Institute of Physics, National Academy of Sciences of Ukraine, Kiev 03028, Ukraine*

We report on a >100-fold enhancement of Raman responses from Bi$_2$Se$_3$ thin films if laser photon energy switches from 2.33 eV (532 nm) to 1.58 eV (785 nm), which is due to direct optical coupling to Dirac surface states (SS) at the resonance energy of ~1.5 eV (a thickness-independent enhancement) and due to nonlinearly excited Dirac plasmon (a thickness-dependent enhancement). Owing to the direct optical coupling, we observed an in-plane phonon mode of hexagonally arranged Se-atoms associated with a continuous network of Dirac SS. This mode revealed a Fano lineshape for films <15 nm thick, resulting from quantum interference between surface phonon and Dirac plasmon states.

## 1. Introduction

Quantum interference (QI) is known to occur if the excited eigenstates in a quantum system are mixtures of the discrete and continuum states [1-9]. This effect manifests itself through the Fano resonance, which is characterized by asymmetric lineshapes in spectra of absorption [1,2], atomic photoionization [3], neutron scattering [4], Raman scattering in heavily-doped Si, carbon nanotubes, and the topological insulator (TI) Bi$_2$Se$_3$ [5-7], electrical transport in single-electron transistors [8], and graphene bilayer infrared absorption [9]. However, Fano resonance observed in the TI Bi$_2$Se$_3$ for a bulk phonon mode, involving out-of-plane vibrations of Bi-Se pairs [7], raises intriguing questions regarding the origin of QI in this material. Because TIs are three-dimensional (3D) materials that are insulating in the bulk (bandgap of Bi$_2$Se$_3$ ~ 0.3 eV) [7], but conductive at the surface due to two-dimensional (2D) Dirac surface states (SS) [10,11], QI in a purely quantum system of Dirac SS would be more likely to involve a surface phonon mode (in-plane vibrations of Se-Se pairs). This type of QI between the discrete phonon and Dirac continuum states is not just of pure scientific interest but also would be of great importance for potential applications of TIs in nanoscale electronics.

In this fast track communication, we report on QI observed in thin films of the TI Bi$_2$Se$_3$, which involves a surface phonon mode of a continuous hexagonal network of Se atoms associated with Dirac SS (H-mode of frequency $\nu_H$) and collective electronic excitation [surface (Dirac) plasmon (SP)] excited within a four-wave mixing (4WM) process [stimulated Raman scattering (SRS)] [12,13]. The latter process deals with nonlinear polarization at difference frequency $\nu_i - \nu_s = \nu_a - \nu_i$ (where $\nu_i$, $\nu_s$, $\nu_a$ denote the frequencies of the incident laser beam, Stokes, and anti-Stokes components, respectively) and is known to appear when the stimulated Raman gain [SRG – the vibrational pumping process occurring in continuous-wave (CW) Raman lasers] [14] at frequency $\nu_i - \nu_s = \nu_H$ exceeds a contrary process of stimulated Raman loss (SRL) at the same frequency of $\nu_a - \nu_i = \nu_H$ [12,13]. Because SRL governs a pump intensity attenuation due to an extra absorption of incident laser light in the free-carrier population of the films, it can be a source of SP excitation and hence QI at frequency $\nu_{SP} \sim \nu_H$. We note that (i) the mechanism of nonlinear SP excitation considered here is similar to that dealing also with 4WM but employing two laser beams [15-17] and (ii) all previous studies of Raman responses from TIs were performed using more energetic laser light (1.96–2.54 eV) [7,18-22] than that used in this study (1.58 eV) and allowed for a direct optical coupling to Dirac SS at the resonance energy of ~1.5 eV [23].

## 2. Sample and experimental setup

Experiments were performed on Bi$_2$Se$_3$ films of different thickness $d$ = 6-40 nm. The films were grown on 0.5 mm Al$_2$O$_3$(0001) substrates by molecular beam epitaxy, with a 10 nm thick MgF$_2$ capping layer to protect against oxidation [24]. Raman measurements were performed using a Renishaw inVia Raman Spectrometer equipped with 532 nm and 785 nm solid-state lasers. The spectra were acquired in a backscattering geometry with 1 cm$^{-1}$ steps and with laser power of 30 mW. A microscope of the spectrometer with a 50× objective lens was used to focus the laser light on the sample surface to a spot size of ~1 µm in diameter. The corresponding power density was $I_L$ = 3.8 MW/cm$^2$. The laser power dependences of Raman responses were measured in the power range of 0.03 - 60 mW ($I_L$ = 3.8 KW/cm$^2$ - 7.6 MW/cm$^2$). The $z(x,x)\bar{z}$ geometry was used, where $z$-axis is directed along a film normal and $x$-axis points to the light polarization. Raman peaks were fitted by Lorentzian/Fano lineshapes to determine their position and linewidth. To determine the free-carrier density, $n_e$, Hall-effect measurements ware used [25].

## 3. Experimental results and discussion

Figures 1(a) and 2(a) show Raman spectra measured with 785 nm laser from the top side of the samples [Fig.1(b), left Inset]. Each spectrum consists of five peaks. Four of them are Raman active 3D phonon modes and their combination: $A_{1g}^1$ (out-of-plane, in-phase) peaked at 73 cm$^{-1}$ (2.2 THz), $E_g^2$ (in-plane, in-phase) peaked at 130 cm$^{-1}$ (3.9 THz), and $A_{1g}^2$ (out-of-plane, out-of-phase) peaked at 173 cm$^{-1}$ (5.2 THz) [Fig.1(c), right Inset] [7,18-22]. The feature peaked at 252 cm$^{-1}$ ($\nu_H$ = 7.56 THz) can be assigned to the 2D H-mode [Fig.1(c), left Inset] since its frequency matches well that of the Se-Se stretching mode for Se$_n$(n > 2)-chains in GeSe glasses [26] and for Se$_8$ rings in monoclinic and amorphous selenium [27]. The hexagonal network of topmost Se-atoms has been proven by second harmonic generation, where six-fold rotational anisotropy was observed [28,29].

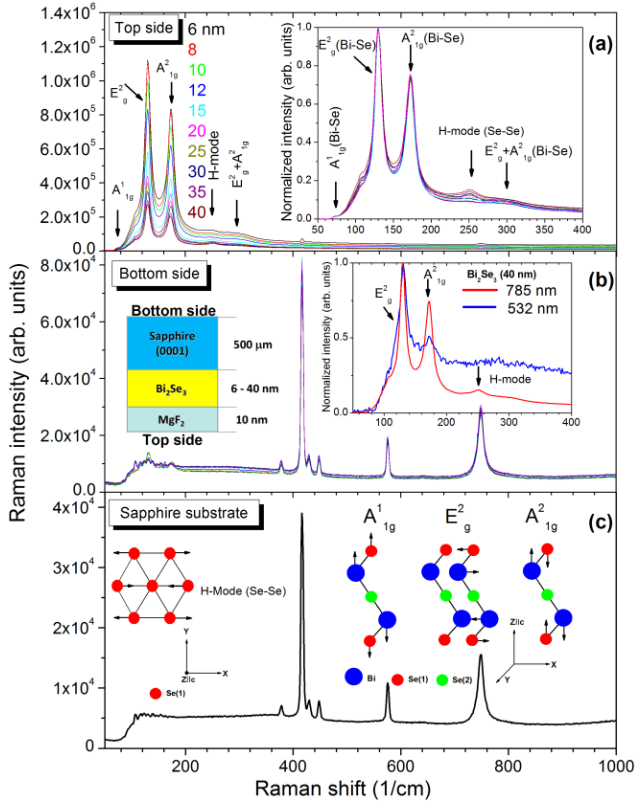

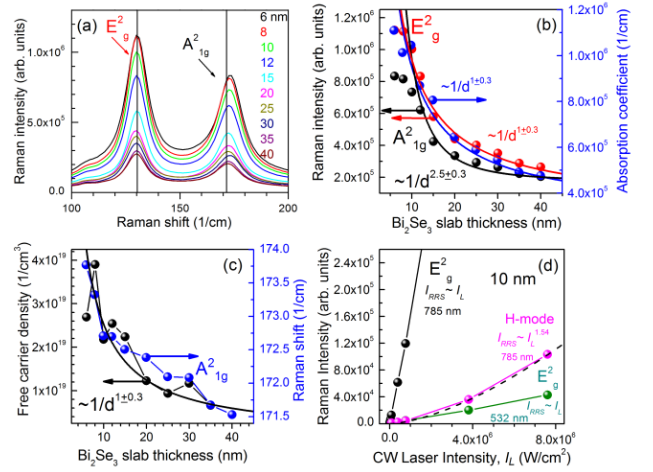

FIG. 1. Raman spectra measured with 785 nm laser light for the top (a) and bottom side (b) of the samples [left inset in (b)] as a function of $Bi_2Se_3$ film thickness. Inset in (a) shows the same Raman spectra but being normalized. Right inset in (b) shows the normalized Raman spectra for the 40 nm film measured with 785 and 532 nm lasers. (c) The Raman spectrum of the 0.5 mm thick sapphire substrate. Insets in (c) present the displacement patterns of phonon modes in $Bi_2Se_3$ films indicated.

FIG. 2. (a) A part of Raman spectra shown in Fig. 1(a). Vertical lines indicate peak positions for the 40 nm film. (b) Thickness dependences of Raman peak intensities for the two 3D phonon modes and absorption coefficient measured with 820 nm (1.51 eV) pulsed laser light [24]. (c) Thickness dependences of $n_e$ and Raman shift of the $A_{1g}^2$ mode. The solid curves in (b) and (c) present a best fit to the data. (d) CW laser power dependences of Raman intensities for the 10 nm film. The dashed line presents the best fit for the H-mode Raman intensity.

Despite the spectral similarity of 3D Raman features measured with 785 nm and 532 nm lasers [Fig. 1(b), Inset], the intensity in the latter case is ~50-fold weaker, resulting in the H-mode peak to be less resolvable. We associate this behavior with resonant Raman scattering (RRS) due to the direct coupling of 785 nm light to Dirac SS [23]. The resulting superlinear (quasi-exponential) laser power dependence of the H-mode Raman intensity ($I_{RRS} \propto I_L^{1.54}$) unambiguously proves SRS to occur [Fig. 2(d)] [12]. In contrast, the 3D phonon mode Raman intensities show the linear-type power dependences for both lasers used [Fig. 2(d)].

The spectra measured from the bottom side of the samples [Fig. 1(b)] are almost identical to that of the sapphire substrate [Fig. 1(c)] [30]. Negligible variations of Raman intensities with decreasing $d$ in the bottom side measured spectra suggest that the corresponding Raman intensity enhancement in the top side measured spectra is due to an additional thickness-dependent source. Correspondingly, the 3D $A_{1g}^2$, $E_g^2$, and $E_g^2 + A_{1g}^2$ phonon modes are enhanced ~4-fold with decreasing $d$, being well fitted as $1/d^{2.5\pm0.3}$ for the $A_{1g}^2$ and $E_g^2 + A_{1g}^2$ modes and as $1/d^{1\pm0.3}$ for the $E_g^2$ mode [Figs. 2(b) and 3(c)]. The 2D H-mode is enhanced only ~2.5-fold as $1/d^{1\pm0.3}$ [Fig. 3(c)]. The overall Raman response enhancement of 3D phonon modes can hence be estimated as >100-fold [Fig. 4(a)].

Despite of the enhancement, all of the spectral trends of 3D Raman responses with decreasing $d$ mainly follow those previously reported [7,18-22]. On the whole, our findings agree well with those measured for $Bi_2Te_3$ nanoplates [31]: the in-plane $E_g^2$ phonon mode remains unshifted, while the out-of-plane $A_{1g}^1$ phonon mode softens (red shift) and $A_{1g}^2$ phonon mode stiffens (blue shift). It should be noted that Raman spectra below 100 cm$^{-1}$ were significantly weakened due to the optical filter of the spectrometer used [Fig. 1(a)]. Consequently, we argue only qualitatively that the $A_{1g}^1$ phonon mode red shifts (not shown) due to phonon softening caused by a reduction of inter-quintuple-layer van-der-Waals forces [7,22]. The blue shift [Fig. 2(a)] suggests that the intra-quintuple-layer interatomic interaction is gradually enhanced with decreasing $d$.

The most important finding is the observation of the H-mode, the linewidth of which non-monotonically depends on $d$, remarkably maximizing for the 10 nm film (Fig. 3). For thinner films (6-12 nm), the H-mode peak is highly asymmetric, suggesting that the Fano resonance takes place [1-9]. The Fano lineshape $I(\nu) = A[f + (\nu - \nu_H)/\Gamma]^2 / 1 + [(\nu - \nu_H)/\Gamma]^2$ allows the electron-phonon coupling strength to be estimated as $1/f = 0.1$, where $f$ is the Fano asymmetry parameter and $\Gamma$ is the linewidth [7]. For thicker films (15-40 nm), the electron-phonon coupling strength decreases to $1/f < 0.1$ and the Fano resonance lineshape reduces to the Lorentzian lineshape. The latter value is consistent with that obtained in angle-resolved photoemission spectroscopy of single $Bi_2Se_3$ crystals ($1/f \sim 0.08$) [32]. In contrast, the Raman linewidth of the $E_g^2 + A_{1g}^2$ phonon mode [as well as the $E_g^2$ and $A_{1g}^2$ modes (not shown explicitly)] has the Lorentzian lineshape over the entire $d$ range [Fig. 3(d)]. We associate this behavior of the H-mode lineshape with strong SP-phonon coupling, at which QI between continuum electronic states of SP and the discrete-energy H-mode phonon occurs. The effect dominants for certain values of $d$, for which $\nu_H \sim \nu_{SP}$, because $n_e$ (and hence $\nu_{SP}$) increases with decreasing $d$ [Fig. 2(c)]. It is worth noting that the $A_{1g}^2$ phonon mode blue shifts in a similar way [Fig. 2(c)]. This behavior can be associated with a 3D-carrier-depletion-induced indirect intersurface coupling [33]. Correspondingly, the increase of $n_e$ results from the upward shift of the conduction band minimum and band unbending when the sum of depletion widths associated with each surface of the film exceeds the film thickness [33,34]. The 3D electron depletion results in an increase of the 2D electron density in Dirac SS, giving rise to the surface-dominated Hall conductivity [35-37]. Furthermore, this kind of indirect intersurface coupling [33,34,38] is expected to unscreen the intra-quintuple-layer interatomic bonding and hence to result in an increase of the stiffness (blue shift) of the 3D $A_{1g}^2$ phonon mode observed.

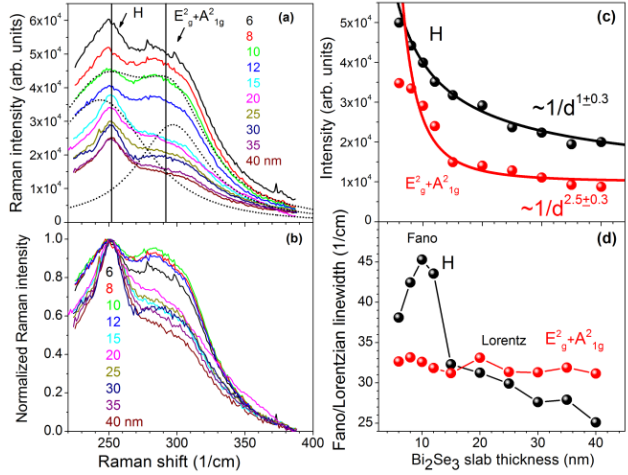

FIG. 3. (a) A part of Raman spectra shown in Fig. 1(a), from which the $A_{1g}^2$ peak Lorentzian tail has been subtracted. The dotted lines show an example of the spectral band decomposition for the 10 nm film. (b) The same spectra but being normalized. (c) and (d) Thickness dependences of Raman intensities and the Fano/Lorentzian linewidths of the phonon modes indicated.

We note that QI does not affect the enhancement trend of the 2D H-mode intensity with decreasing $d$ [Fig. 3(c)]. In contrast, the enhancement of the 3D phonon modes tends to stabilize for the thinnest films [Figs. 2(b) and 3(c)]. This behavior can be seen more clearly through the ratio between Raman intensities measured with 785 nm and 532 nm lasers [Fig. 4(a)]. Figures 2(b), 3(c), and 4(a) also indicate that the 3D in-plane phonon mode ($E_g^2$) and the 2D H-mode reveal similar trends with decreasing $d$, pointing to joint displacements of Se-atoms involved [Fig.1(c), Insets]. Concurrently, the different $d$ trends of in-plane and out-of-plane phonon modes point to two different sources of the Raman response enhancement.

To treat the thickness-dependent part of Raman response enhancement in $Bi_2Se_3$ thin films, we applied the model developed for SPs in electrostatically coupled graphene bilayers [39-43], using the corresponding dielectric constants for the $Al_2O_3/Bi_2Se_3/MgF_2$ stack and taking into account the 3D-carrier-depletion-induced indirect intersurface coupling [33,34,38]. We assumed that the enhancement is due to the electric field associated with both longitudinal (TM) and transverse (TE) SPs of two modes (optical SP mode with $\sqrt{q}$-dispersion and the SS-separation($d$)-dependent acoustic SP mode with linear $q$-dispersion, where $q$ is the SP wavevector) [39-43]. Because the frequency of TM optical SP is expected to be much lower than that of TM acoustic SP [39], its contribution to the dynamics discussed is negligible. Alternatively, the TE optical SP mode is known to dominate over the small $d$ range used [39,40]. We therefore considered only two SP modes for $Bi_2Se_3$ thin films: TE optical (in-phase) SP and TM acoustic (out-of-phase) SP modes, for which in-phase and out-of-phase oscillations are purely charge- and spin-like, respectively, due to the spin-charge separation effect [Figs. 4(c) and 4(d)] [40]. Correspondingly, QI occurs between TM acoustic SP and the H-mode due to the spin-momentum locking in Dirac SS and the resulting co-direction of the spin SP electric field oscillations and the H-mode wavevector [44]. Correspondingly, the excitation of the two different SP modes is responsible for the different enhancement trends of in-plane and out-of-plane phonon modes with decreasing $d$. The further stabilization of the 3D mode enhancement for the thinnest films [Fig. 4(a)] is likely due to the 3D-carrier depletion effect [33], which also does not affect the in-plane H-mode enhancement through the charge-less TM acoustic SP [Fig. 3(a) and 3(c)] [40].

The RRS intensity can be given in terms of the complex dielectric function $\varepsilon(\nu_i)$ [45,46],

$$I_{RRS} \propto \left|\partial\varepsilon(\nu_i)/\partial h\nu_{ph}\right|^2 \quad . \quad (1)$$

where the derivative with respect to the phonon energy $h\nu_{ph}$ in the vicinity of resonant electronic transition in Dirac SS (~1.5 eV) [23] is applied similarly to that used for RRS in the vicinity of absorption continua [45]. The approximation is valid because SRS involves a nonlinear polarization in free-carrier population at low frequency $\nu_{ph}$. Also, this approach implies that the absorption coefficient at the incident frequency is governed by the sum of the 3D interband electronic transitions, the electronic transitions in 2D Dirac SS, and the intraband free-carrier transitions (Drude absorption) at the difference frequency $\nu_a - \nu_i = \nu_{ph}$ [Fig. 4(b)],

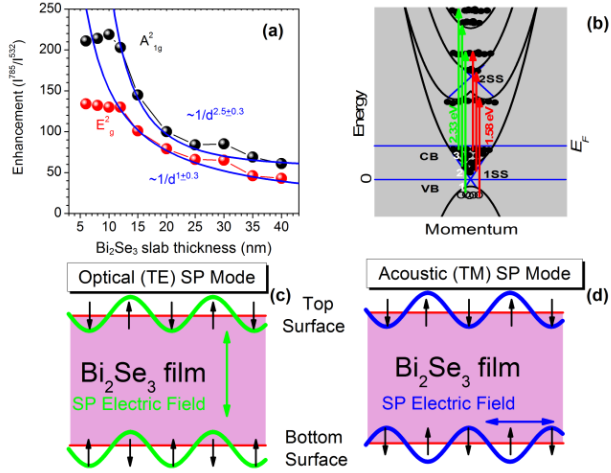

FIG. 4. (a) Thickness dependences of 3D phonon modes enhancement ($I^{785}/I^{532}$). The solid blue curves present a best fit to the data. (b) Electronic transitions initiated by 785 nm (1.58 eV) and 532 nm (2.33 eV) light are shown by the corresponding color arrows over the Bi$_2$Se$_3$ band structure. The position of Fermi energy ($E_F$) is shown. 1,2,3 denote electronic transitions started from the valence band maximum, occupied Dirac SS, and below the Fermi level, respectively. (c) and (d) Schematic presentation of two SP modes in the intersurface-coupled Bi$_2$Se$_3$ films. The spin polarization and SP electric field oscillations are shown.

$$\alpha(\nu_i) \propto \text{Im}[\varepsilon(\nu_i)] \propto \alpha_{3D}(\nu_i) + \alpha_{2D}(\nu_i) + \alpha^e(\nu_{ph}). \quad (2)$$

The first two terms are assumed to be constant and therefore the second term is responsible for the thickness independent ~50-fold enhancement of Raman responses, whereas the last contribution provides the thickness-dependent dynamics,

$$I_{RRS} \propto \left|\partial \alpha^e(\nu_{ph})/\partial h \nu_{ph}\right|^2. \quad (3)$$

This equation implies that SRL governs an extra absorption of incident light [12,13] in the free-carrier population and hence controls the efficiency of SP excitation at the difference frequency $\nu_a - \nu_i = \nu_{ph}$:

$$\alpha^e(\nu_{ph}) \propto \text{Im}[\varepsilon^e(\nu_{ph})] \propto \text{Im}[\chi^{(3)}_{xxxx}(-\nu_i; \nu_a, \nu_i, -\nu_a)]$$
$$\propto A/\{[(\nu_a - \nu_i) - \nu_{ph}]^2 + \gamma_e^2\} \quad (4)$$

where $\varepsilon^e(\nu_{ph})$, $\chi^{(3)}_{xxxx}$, $A$, and $\gamma_e$ are the free-electron part of the complex dielectric function, the Raman in-plain third-order susceptibility, the effective amplitude, and the electron-electron scattering rate, respectively [13]. This absorption process in free-carrier population is also consistent with an increase of the total absorption coefficient of Bi$_2$Se$_3$ films with decreasing $d$ experimentally observed using 1.51 eV pulsed light [Fig. 2(b)] [24].

The excitation of SP is expected to occur if $\nu_{SP}$ for a certain $n_e$ (for a certain $d$) is in the vicinity of $\nu_H$ since momentum matching in this case is obeyed automatically. The dispersion for the two SP modes discussed [39-43] can be expressed as

$$\nu^{TM}_{SP}(q) = (1/2\pi)2\sqrt{n_{2D}e^2 dq^2/(1+\varepsilon_s)\varepsilon_0 m^*}, \quad (5)$$

$$\nu^{TE}_{SP}(q) = (1/2\pi)\sqrt{n_{2D}e^2 q/(1+\varepsilon_s)\varepsilon_0 m^*}, \quad (6)$$

where the spin- and valley degeneracies $g_s = g_v = 1$ were omitted [40], $n_{2D}$ is the 2D carrier density, $e$ is electron charge, $m^* = 0.13 m_0$ is the electron effective mass with $m_0$ being the free-electron mass, $\varepsilon_0$ is the permittivity of free space, and $\varepsilon_s$ is the effective low-frequency dielectric constant [33]. We note that despite the estimation of 2D carrier density in thin films of the TI Bi$_2$Se$_3$ is not trivial [34], in the simple capacitor-like model of 3D-carrier-depletion-induced intersurface coupling [33] $n_{2D}$ and $q$ can be approximated as $n_e^{2/3}$ and $n_e^{1/3}$, respectively [47]. The latter statement can be confirmed by taking into account the ratio of SP frequencies presented in Eqs. (5) and (6) as $\nu^{TM}_{SP}(q)/\nu^{TE}_{SP}(q) = 2\sqrt{dq}$. Assuming that $\nu_H = \nu^{TM}_{SP} = 7.56$ THz and using free-carrier density in the 12 nm film ($n_e \sim 2.54 \times 10^{19}$ cm$^{-3}$), for example, the frequency of TE optical (in-phase) SP that provides the enhancement of the out-of-plane phonon modes can be estimated as $\nu^{TE}_{SP} = 2.01$ THz, which matches well the frequency of $A^1_{1g}$ (out-of-plane, in-phase) Raman mode (2.2 THz).

The Drude absorption coefficient can also be rewritten in a traditional form [45]:

$$\alpha^e(\nu_{ph}) = \varepsilon_s \nu^2_{SP} \gamma_e/[cn_r(\nu^2_{ph} + \gamma^2_e)], \quad (7)$$

where $c$ is speed of light, $\gamma_e^{-1} \leq 50$ fs [48], i.e. $\gamma_e \geq 20$ THz, and $n_r$ is the real part of refractive index, which is approximately independent of $\nu_{ph}$ [45]. Using Eq. (7) in Eq. (3) and assuming that $\nu_{ph} = \nu_H = \nu_{SP} \ll \gamma_e$ yields

$$I_{RRS} \propto \nu^6_{SP}. \quad (8)$$

Taking $\nu_{SP}$ in the form of Eqs. (5) and (6), the proportionality of Eq. (8) can be rewritten as

$$I^{TM}_{RRS} \propto n^3_{2D} d^3 q^6 \propto n^2_e d^3 n^2_e \propto n^4_e d^3, \quad (9)$$

$$I^{TE}_{RRS} \propto n^3_{2D} q^3 \propto n^2_e n_e \propto n^3_e. \quad (10)$$

To verify whether these dependences are relevant to the experimental observations, we approximate $n_e \propto 1/d$ [Fig. 2(c)]. According to Eq. (7), the increase of $\alpha^e$ with decreasing $d$ is dominated by the excitation of TE optical SP (increases as $1/d$) as compared to the excitation of TM acoustic SP (increases as $1/d^{1/3}$) [Fig. 2(b)] [24]. Consistently, the enhancement of Raman responses associated with excitations of TM acoustic SP and TE optical SP vary with decreasing $d$ as $1/d$ and $1/d^3$, respectively [Eqs. (9) and (10)], i.e. agrees well with observations presented in Figs. 2(b), 3(c), and 4(a).

### 4. Conclusion

In summary, we observed a >100-fold enhancement of Raman responses from thin films of the TI Bi$_2$Se$_3$ if the photon energy of laser light matches the energy of resonant electronic transition in Dirac SS. The in-plane phonon mode associated with a continuous network of hexagonally arranged Dirac SS was observed. This mode revealed a thickness-dependent Fano resonance, which results from QI between H-mode phonon and Dirac plasmon states. Our observations are in good agreement with the model developed for SPs in electrostatically coupled graphene bilayers.

**Acknowledgment**

This work was supported by a Research Challenge Grant from the West Virginia Higher Education Policy Commission (HEPC.dsr.12.29). The work was performed using the West Virginia University Shared Research Facilities.